%% file: kernel.tex
\documentclass[]{elsarticle}

\usepackage{lineno,hyperref}
\modulolinenumbers[5]

\journal{0720}

\biboptions{authoryear}

\begin{document}

\begin{frontmatter}

\title{Kernel density decomposition with an application to the social cost of carbon\tnoteref{mytitlenote}}
\tnotetext[mytitlenote]{The replication package and generic code are on \href{https://github.com/rtol/KernelDecomposition}{GitHub}.
}

\author[label2,label3,label4,label5,label6,label7]{Richard S.J. Tol\corref{cor1}\fnref{label8}}
\address[label2]{Department of Economics, University of Sussex, Falmer, United Kingdom}
\address[label3]{Institute for Environmental Studies, Vrije Universiteit, Amsterdam, The Netherlands}
\address[label4]{Department of Spatial Economics, Vrije Universiteit, Amsterdam, The Netherlands}
\address[label5]{Tinbergen Institute, Amsterdam, The Netherlands}
\address[label6]{CESifo, Munich, Germany}
\address[label7]{Payne Institute for Earth Resources, Colorado School of Mines, Golden, CO, USA}

\cortext[cor1]{Jubilee Building, BN1 9SL, UK}
\fntext[label7]{Earlier versions of this paper were presented at the University of Sussex and the University of Surrey.}

\ead{r.tol@sussex.ac.uk}
\ead[url]{http://www.ae-info.org/ae/Member/Tol\_Richard}

\begin{abstract}
A kernel density is an aggregate of kernel functions, which are itself densities and could be kernel densities. This is used to decompose a kernel into its constituent parts. Pearson's test for equality of proportions is applied to quantiles to test whether the component distributions differ from one another. The proposed methods are illustrated with a meta-analysis of the social cost of carbon. Different discount rates lead to significantly different Pigou taxes, but not different growth rates. Estimates have not varied over time. Different authors have contributed different estimates, but these differences are insignificant. Kernel decomposition can be applied in many other fields with discrete explanatory variables. 
\end{abstract}


\end{frontmatter}

%\linenumbers

\section{Introduction}
Everything about climate change is uncertain. Many estimates of the social cost of carbon have been published, and kernel densities have been used to visualize the uncertainty about the Pigou tax \citep{Tol2018}. I here propose a new method to decompose that uncertainty into discrete components, and a new statistical test for whether the components differ from one another.

Kernel densities are a useful tool to describe univariate data \citep{Takezawa2005}. Simple kernel regression is helpful for specifying the relationship between two variables \citep{Altman1992}, and kernel quantile regression can be used to show this relationship across the distribution \citep{Yu1998}. However, these methods are not suitable if the explanatory variable is categorical. The method proposed here works well for categorical data, and shows both central tendency and spread.

This paper extends previous meta-analyses of the social cost of carbon \citep{Tol2005, Tol2018}. The kernel function and bandwidth are the same as in the 2018 paper (see \ref{app:kernel}), but there are now many more estimates (see \ref{app:estimates}). My earlier papers used sample splits rather than decomposition and did not conduct statistical tests for differences. \citet{Wang2019} report a meta-analysis of 548 estimates of the social cost of carbon. I have 2786. They focus on the central tendency, while I consider the entire distribution. They assume linearity and normality. I do not.

The paper proceeds as follows. Section \ref{sc:decomp} presents the decomposition method, and Section \ref{sc:test} the appropriate hypothesis test. Section \ref{sc:apply} applies these methods to estimates of the social cost of carbon and its growth rate. Section \ref{sc:conclude} concludes.

\section{Kernel density decomposition}
\label{sc:decomp}
A \textit{kernel density} is defined as
\begin{equation}
\label{eq:kernel}
    f(x) = \frac{1}{nh} \sum_{i=1}^n K \left ( \frac{x-x_i}{h} \right )
\end{equation}
where $x_i$ are a series of observations, $h$ is the bandwidth, and $K$ is the \textit{kernel function}. The kernel function is conventionally assumed to be a (\textit{i}) \textbf{non-negative} (\textit{ii}) \textbf{symmetric} function that (\textit{iii}) \textbf{integrates to one}, with (\textit{iv}) \textbf{zero mean} and (\textit{v}) \textbf{finite variance} \citep{Takezawa2005}. That is, any standardized symmetric probability density function can serve as a kernel function, and the Normal density is indeed a common choice.

Conventions are just that. As long as the kernel function is non-negative\footnote{The assumption of non-negativity is relaxed for bias reduction \citep{Jones1997}, a topic unrelated to the current paper.} and integrates to one, an appropriately weighted sum of kernel functions is non-negative and integrates to one\textemdash such a sum is a probability density function. Of course, if the kernel function is \textbf{asymmetric}, centralization needs to be carefully considered\textemdash is $x_i$ the mean, median or mode of $K$?

A kernel density can be seen as a mixture \citep{Makov2001, McLachlan2001}.\footnote{\citet{Quetelet1846, Quetelet1852} was the first to note that the weighted sum of densities is a density, \citet{Pearson1894} the first to apply this insight.} This reinterpretation opens a route to decomposition. We can construct the kernel density of any subset of $x_i$. The weighted sum of the kernel densities of the subsets is a kernel density.

Indeed, with the right weights and bandwidths, the weighted sum of the kernel densities of subsets of the data is identical to the kernel density of the whole data set. To see this, partition the observations into $m$ subsets of length $m_j$ with $\sum_j m_j = n$, as $x_1, ... x_{m_1}, x_{m_1+1}, ..., x_{m_1+m_2}, x_{m_1+m_2+1}, ..., x_n$. Then
\begin{equation}
\label{eq:decomp}
    f(x) = \sum_{j=1}^m \frac{m_j}{n} \frac{1}{m_j h} \sum_{i=\sum_{k=1}^{j-1} m_k +1}^{\sum_{k=1}^{j} m_k} K \left ( \frac{x-x_i}{h} \right ) =: \sum_{j=1}^m \frac{m_j}{n} f_j(x)
\end{equation}
This is identical to Equation (\ref{eq:kernel}). Moreover, each of the components $f_j$ of the composite kernel density $f$ is itself a kernel density.

I doubt I am the first to notice this, but as far as I know I am the first to write it up.\footnote{I searched Scopus for "kernel decomposition", "composite kernel" and "kernel mixture". \citet{Cunningham1994} show that a spectogram is the weighted sum of spectograms. \citet{Szymkowiak2006} decompose kernels by introducing a conditional variable (much like I do) but use this for spectral clustering rather than decomposition. \citet{Szafranski2010} fit alternative kernel densities to the same data, and construct a composite kernel density using model fit as weights; see \citet{Kloft2011} for a discussion of the appropriate weights.}

Note that kernel decomposition works with any set of weights that add to one; and with any kernel function or bandwidth for the subsets:
\begin{equation}
\label{eq:mixture}
    f(x) = \sum_{j=1}^m \frac{w_j}{m_j h_j} \sum_{i=\sum_{k=1}^{j-1} m_k +1}^{\sum_{k=1}^{j} m_k} K_j \left ( \frac{x-x_i}{h_j} \right ) =: \sum_{j=1}^m w_j f_j(x;h_j)
\end{equation}
In this case, the composite kernel density is not be the same as the kernel density fitted to the complete data set. While it may be hard to argue in favour of different kernel functions $K_j$ for different subsets of the data, meaningfully different subsets of the data would have different spreads and hence bandwidths $h_j$.

The mathematics work for \textit{any} partitioning of the observations, but decomposing a kernel density in this manner makes as much sense as the partitioning. Below, I take a large sample of estimates of the social cost of carbon and decompose its kernel density in three ways: by author, by year of publication, and by discount rate. As the partitioning is intuitive, so is the decomposition.

Three hypotheses are tested: Do different discount rates imply different climate policy? The expected answer is yes. Do estimates change over time? In other words, are we learning? The answer is let's hope so. Do different researchers reach different conclusions? In other words, do subjective judgements affect the estimates? The answer is let's hope not.

The decomposition of the kernel density by author opens another interpretation: Vote-counting \citep{Laplace1814}. Different experts have published different estimates of the social cost of carbon. These can be seen as votes for a particular Pigou tax. But as the experts are uncertain, they have voted for a central value and a spread. The kernel function is a vote, the kernel density adds those votes.\footnote{Note the difference with Bayesian updating, which multiplies rather than adds probabilities.}

Different experts have cast different numbers of votes. Below, I count one paper as one vote. One can also argue that it should be one vote per expert, or that papers should be weighted by citations, journal prestige, or author pedigree. Composite kernel densities naturally allow for this, but it is a dangerous route to travel in this case.

\section{Inference}
\label{sc:test}
Equation (\ref{eq:mixture}) holds that the kernel density $f(x)$ is composed of $m$ kernel densities $f_j(x)$ with weight $m_j / n$. For each interval $\underline x < x < \bar x$, we can test whether the shares of the component densities equal the overall weights, using the Equality of Proportions test by \citet{Pearson1900}. Such a test reveals whether a component density disproportionally contributes to, say, the left tail of the composite density.

If the intervals correspond to $p$ percentiles of the composite distribution, the test statistic is
\begin{equation}
\label{eq:pearson}
    \chi_{(m-1)(p-1)}^2 = n \sum_{k=0}^p \sum_{j=1}^m \frac{\left( \int_{P_k}^{P_{k+1}} f_j(x) \mathrm{d} x - \frac{m_j}{n} \right )^2}{\frac{m_j}{n}}
\end{equation}
Note that the test only works if there are two components or more, $m \geq 2$. If not, there would be nothing to compare. Note also that the distribution needs to be split in two quantiles or more, $p \geq 2$. This is because each component density adds up to its weight $m_j / n$ by construction.

Matlab codes to construct and decompose a kernel and test the decomposition are on \href{https://github.com/rtol/KernelDecomposition}{GitHub}.

\section{Application}
\label{sc:apply}

\subsection{The social cost of carbon}
The social cost of carbon is the damage done, at the margin, by emitting more carbon dioxide into the atmosphere. If evaluated along the optimal emissions trajectory, the social cost of carbon equals the \citet{Pigou1920} tax that internalizes the externality and restores the \citet{Pareto1906} optimum. Climate change features high on the political and public agenda. Estimating the social cost of carbon is an intellectual challenge, requiring the projection of emissions, concentrations and climates over a long period of time, the estimation the effects of climate change, the valuation of a wide variety of impacts, the aggregation of impacts over people with a huge range of living standards, the consideration of large uncertainties and ambiguous knowledge, and the discounting of welfare in the distant future and over multiple generations \citep{Tol2018}.

It is therefore no surprise that there is a large literature on the social cost of carbon spanning four decades, from \citet{Nordhaus1982} to \citet{Okullo2020}. I count 2786 estimates in 148 papers. These are estimates for the social cost of carbon of carbon dioxide emitted in the recent past. 54 papers published estimates of the social cost of carbon at two or more points in time, for a total of 648 estimates of the \textit{growth rate} of the social cost of carbon.\footnote{The carbon tax should increase over time until emissions are so low that the marginal impacts of climate change start to fall.}

\subsection{Discount rate}
\begin{figure}
    \centering
    \includegraphics[width=\textwidth]{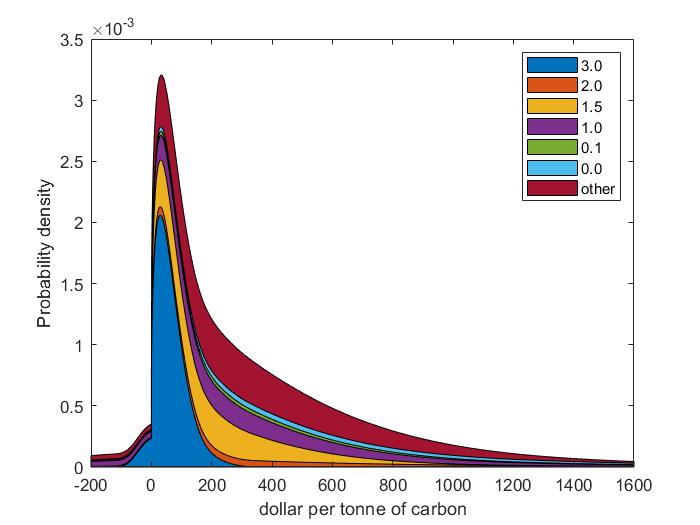}
    \caption{Kernel density of the social cost of carbon and its composition by discount rate.}
    \label{fig:discount}
\end{figure}

Figure \ref{fig:discount} decomposes the kernel density of the social cost of carbon into its components by pure rate of time preference used. Three choices of time preference are most common: 3.0\%, 1.5\% and 1.0\%.\footnote{"Other" refers to a range of numbers and methods, but mostly constant consumption rates.} As one would expect, the lower discount rates contribute more to the right tail of the distribution.

Table \ref{tab:discount} shows the contributions of estimates of the social cost of carbon using a particular pure rate of time preference to the overall kernel density (denoted "null") as well as to the five quintiles of that density (denoted Q1-5). The null hypothesis that all shares are equal is firmly rejected; $\chi^2_{24} = 98.6; p = 0.000$.

\subsection{Author}
I split the sample into estimates by those who have published five papers or more (i.e., Christopher W. Hope, William D. Nordhaus, Frederick  van der Ploeg, Richard S.J. Tol) and others.

\begin{figure}
    \centering
    \includegraphics[width=\textwidth]{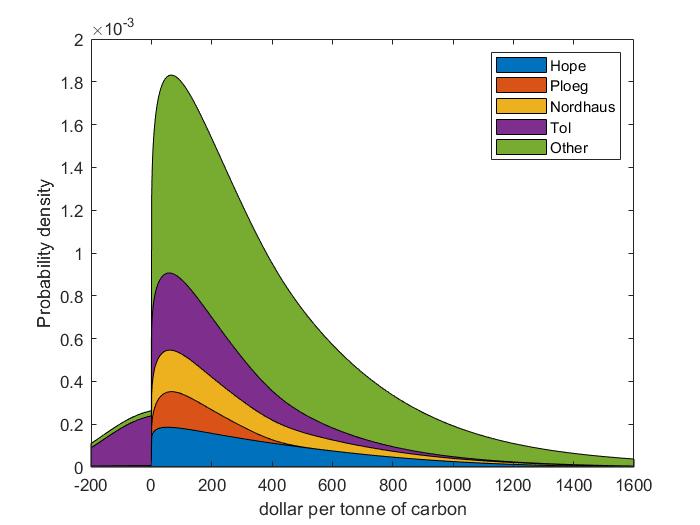}
    \caption{Kernel density of the social cost of carbon and its composition by author.}
    \label{fig:author}
\end{figure}

Figure \ref{fig:author} decomposes the kernel density by author. Of the named authors, estimates by van der Ploeg are the narrowest, Tol contributes most to the left tail, and Hope to the right tail.

Table \ref{tab:author} shows the contributions of estimates of the social cost of carbon published by a particular author to the overall kernel density and its quintiles. Although there are patterns in figure and table, the quintile shares are statistically indistinguishable from the overall shares; $\chi^2_{16} = 19.2; p = 0.260$.

\subsection{Year of publication}
I split the sample into five periods. The key events are the Second Assessment Report of the Intergovernmental Panel on Climate Change, and the first IPCC report to discuss the economic impacts \citep{Pearce1996}, the Third Assessment Report of the IPCC \citep{Smith2001}, the Stern Review \citep{Stern2006}, and the Obama update of the social cost of carbon \citep{IAWGSCC2013}.

\begin{figure}
    \centering
    \includegraphics[width=\textwidth]{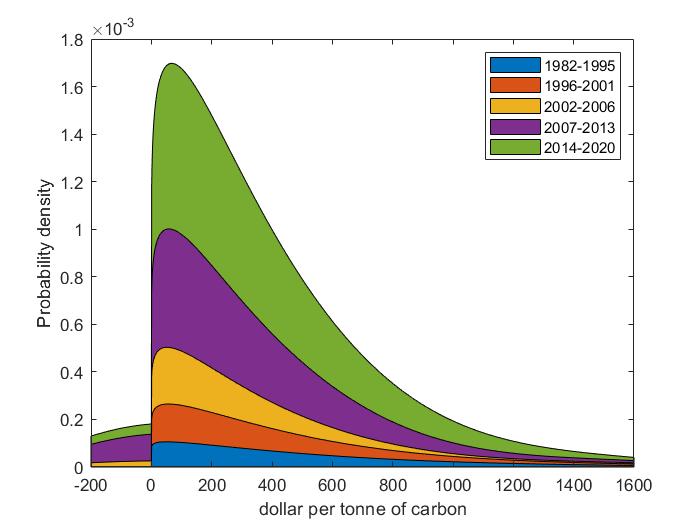}
    \caption{Kernel density of the social cost of carbon and its composition by publication period.}
    \label{fig:period}
\end{figure}

Figure \ref{fig:period} shows the decomposition by publication period. The earlier studies excluded negative estimates, but other patterns are not obvious.

Table \ref{tab:period} shows the contributions of estimates of the social cost of carbon published in a particular period to the overall kernel density and its quintiles. The quintile shares are statistically indistinguishable from the overall shares; $\chi^2_{16} = 4.14; p = 0.999$.

\subsection{The growth rate of the social cost of carbon}
\begin{figure}
    \centering
    \includegraphics[width=\textwidth]{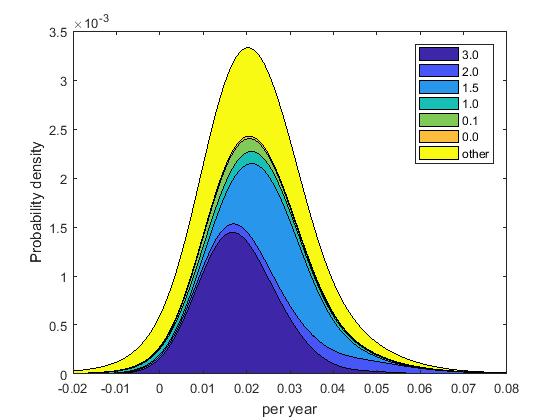}
    \caption{Kernel density of the growth rate of the social cost of carbon and its composition by discount rate.}
    \label{fig:growth}
\end{figure}

Figure \ref{fig:growth} shows the kernel density of the growth rate of the social cost of carbon, decomposed for the pure rate of time preference. The density is symmetric for a 3.0\% utility discount rate. However, for discount rates of 1.5\% and 2.0\%, little probability mass is added to the left tail, and a lot to the right tail.

Table \ref{tab:growth} shows the shares by quintile of the kernel density. The same pattern is seen as in the graph, but Pearson's test for the equality of proportions does not reject the null hypothesis that the component densities are equal to the composite one; $\chi^2_{24} = 10.6; p = 0.992$.

\section{Discussion and conclusion}
\label{sc:conclude}
I present a method to decompose kernel densities and statistically test whether the components differ from the composite. I illustrate the proposed method with a meta-analysis of the social cost of carbon and its growth rate. As expected, a lower discount rate implies a statistically significantly higher Pigou tax; its growth rate is not affected by the discount rate. Earlier estimates of the social cost of carbon are not different from later estimates. There appears to be no learning over time. The null hypothesis that different authors publish similar estimates cannot be rejected. Published estimates are not marked by subjectivity.

The proposed method can be applied to any issue with categorical explanatory variables, such as the wages of men and women, energy use of home owners and renters, commuting times by public and private transport, exam scores by ethnicity, or projected corona virus deaths by political affiliation.

The method presented here considers a univariate kernel density and its dependence on a single categorical variable. Generalization to bi- and trivariate kernel densities is immediate, although visualization would be a challenge. If a kernel density can be used as kernel function to form a composite kernel density, then the composite kernel density can of course also be used as a kernel function\textemdash and so on. That is, although I present a single nest for a single categorical explanatory variable, deeper nests are possible if so desired.

\bibliography{kernel}

\appendix

\section{New estimates of the social cost of carbon}
\label{app:estimates}
The previous meta-analysis of the social cost of carbon \citep{Tol2018} was extended with estimates reported in \citet{Anthoff2019}, \citet{Bretschger2019}, \citet{Budolfson2017}, \citet{Daniel2019}, \citet{Dayaratna2020}, \citet{Ekholm2018}, \citet{Faulwasser2018}, \citet{Golub2017}, \citet{Guivarch2018}, \citet{Hafeez2017}, \citet{Hansel2018}, \citet{Kotchen2018}, \citet{Moore2017}, \citet{Nordhaus2015}, \citet{Okullo2020}, \citet{Ricke2018}, \citet{Scovronick2017}, \citet{Tol2019}, \citet{Yang2018} and \citet{Zhen2018}. The Budolfson and Faulwasser estimates were digitized from graphs.

\citet{Glanemann2020} do not report a carbon tax, \citet{Zhen2019} report the \textit{relative} social cost of carbon, \citet{Kelleher2019} \textit{relative changes} in the social cost of carbon, \citet{Ploeg2019} the \textit{steady state} social cost of carbon, and \citet{Pindyck2017, Pindyck2019} the \textit{average} social cost of carbon.

All data to reproduce this paper are on \href{https://github.com/rtol/KernelDecomposition}{GitHub}.

\section{Bandwidth and kernel function}
\label{app:kernel}
\begin{figure}
    \centering
    \includegraphics[width=\textwidth]{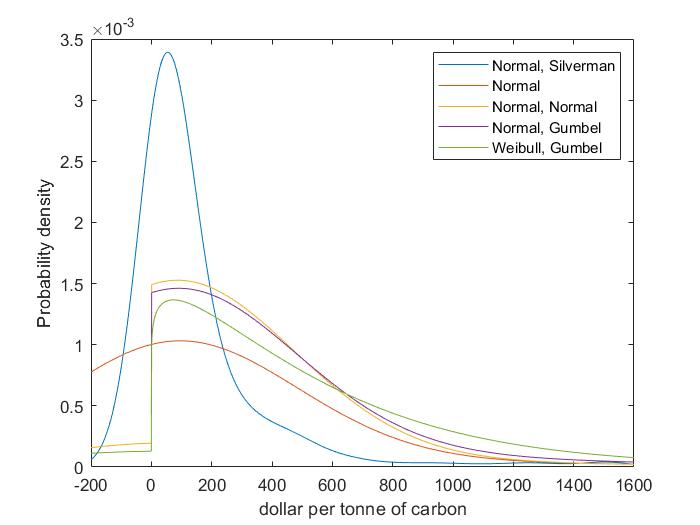}
    \caption{Kernel density of the social cost of carbon for alternative kernel functions and bandwidths.}
    \label{fig:kernel}
\end{figure}

The choice of kernel function and bandwidth is key to any kernel density estimate, as illustrated in Figure \ref{fig:kernel}. Kernel and bandwidth should be chosen such to reflect the nature of the data. In this case, the uncertainty about the social cost of carbon is large and right-skewed. Furthermore, the social cost of carbon is, most researchers argue, a cost and not a benefit.

A conventional choice would be to use a Normal kernel function, with a bandwidth according to the Silverman rule, that is, 1.06 times the sample standard deviation divided by the number of observations raised to the power one-fifth. Figure \ref{fig:kernel} reveals two problems with this approach: The right tail is thin, and a large probability mass is assigned to negative social costs of carbon. If the bandwidth equals the sample standard deviation to reflect the wide uncertainty, the right tail appropriately thickens but the probability of a Pigou subsidy on greenhouse gas emissions increases too.

Many of the published estimates of the social cost of carbon are based on an impact function that excludes benefits of climate change. Honouring that, I assign a knotted Normal kernel function to these observations, with the mode as its central tendency. The probability of a negative social cost of carbon falls.

The studies that report the possibility of a negative social cost of carbon nonetheless argue in favour of a positive one. A symmetric Normal kernel does not reflect that. I therefore replace it with a Gumbel kernel, which is defined on the real line but right-skewed. Again, I use the mode as its central tendency. This thickens the right tail and thins the left tail.

A knotted Normal kernel is not only peculiar near zero but it also has a thin tail. I therefore replace it with a Weibull kernel, which is defined on the positive real line, near zero near zero, and right-skewed. I use the mode as its central tendency. The right tail of the kernel distribution thickens again. The Weibull-Gumbel kernel distribution is the default used here.

Code and data to reproduce this paper are on \href{https://github.com/rtol/KernelDecomposition}{GitHub}.

\newpage \section{Additional results}
\label{app:results}

\begin{table}[h]
    \centering
    \include{discount}
    \caption{Observed and hypothesized contribution to the kernel density by quintile and pure rate of time preference.}
    \label{tab:discount}
\end{table}

\begin{table}[h]
    \centering
    \include{author}
    \caption{Observed and hypothesized contribution to the kernel density by quintile and author.}
    \label{tab:author}
\end{table}

\begin{table}[h]
    \centering
    \include{period}
    \caption{Observed and hypothesized contribution to the kernel density by quintile and publication period.}
    \label{tab:period}
\end{table}

\begin{table}[h]
    \centering
    \include{growth}
    \caption{Observed and hypothesized contribution to the kernel density of the growth rate of the social cost of carbon by quintile and pure rate of time preference.}
    \label{tab:growth}
\end{table}

\end{document}

%% file: discount.tex
\begin{tabular}{|l|c|c|c|c|c|c|c|}
\hline
&\textbf{3.0}&\textbf{2.0}&\textbf{1.5}&\textbf{1.0}&\textbf{0.1}&\textbf{0.0}&\textbf{other}\\\hline
\textbf{Q1}&0.1677&0.0066&0.0368&0.0354&0.0050&0.0074&0.0544\\\hline
\textbf{Q2}&0.0520&0.0085&0.0470&0.0271&0.0045&0.0061&0.0567\\\hline
\textbf{Q3}&0.0028&0.0088&0.0399&0.0284&0.0051&0.0077&0.0614\\\hline
\textbf{Q4}&0.0000&0.0099&0.0289&0.0295&0.0060&0.0107&0.0680\\\hline
\textbf{Q5}&0.0000&0.0168&0.0119&0.0347&0.0096&0.0268&0.0779\\\hline
\textbf{Null}&0.0445&0.0101&0.0329&0.0310&0.0060&0.0117&0.0637\\\hline
\end{tabular}

%% file: author.tex
\begin{tabular}{|l|c|c|c|c|c|}
\hline
&\textbf{Hope}&\textbf{Nordhaus}&\textbf{Ploeg}&\textbf{Tol}&\textbf{Other}\\\hline
\textbf{Q1}&0.0184&0.0135&0.0175&0.0722&0.0900\\\hline
\textbf{Q2}&0.0218&0.0173&0.0217&0.0403&0.1140\\\hline
\textbf{Q3}&0.0218&0.0092&0.0196&0.0330&0.1172\\\hline
\textbf{Q4}&0.0233&0.0009&0.0174&0.0225&0.1223\\\hline
\textbf{Q5}&0.0220&0.0000&0.0146&0.0097&0.1397\\\hline
\textbf{Null}&0.0214&0.0082&0.0182&0.0355&0.1166\\\hline
\end{tabular}

%% file: period.tex
\begin{tabular}{|l|c|c|c|c|c|}
\hline
&\textbf{1982-1995}&\textbf{1996-2001}&\textbf{2002-2006}&\textbf{2007-2013}&\textbf{2014-2020}\\\hline
\textbf{Q1}&0.0095&0.0142&0.0284&0.0758&0.0754\\\hline
\textbf{Q2}&0.0126&0.0190&0.0267&0.0595&0.0862\\\hline
\textbf{Q3}&0.0130&0.0189&0.0239&0.0586&0.0874\\\hline
\textbf{Q4}&0.0143&0.0191&0.0204&0.0569&0.0879\\\hline
\textbf{Q5}&0.0219&0.0229&0.0140&0.0468&0.0867\\\hline
\textbf{Null}&0.0143&0.0188&0.0227&0.0595&0.0847\\\hline
\end{tabular}

%% file: growth.tex
\begin{tabular}{|l|c|c|c|c|c|c|c|}
\hline
&\textbf{3.0}&\textbf{2.0}&\textbf{1.5}&\textbf{1.0}&\textbf{0.1}&\textbf{0.0}&\textbf{other}\\\hline
\textbf{Q1}&0.0628&0.0054&0.0056&0.0125&0.0053&0.0029&0.0772\\\hline
\textbf{Q2}&0.1029&0.0057&0.0227&0.0091&0.0104&0.0019&0.0644\\\hline
\textbf{Q3}&0.0858&0.0072&0.0527&0.0085&0.0084&0.0017&0.0609\\\hline
\textbf{Q4}&0.0540&0.0108&0.0726&0.0083&0.0045&0.0017&0.0582\\\hline
\textbf{Q5}&0.0207&0.0339&0.0484&0.0087&0.0028&0.0028&0.0588\\\hline
\textbf{Null}&0.0652&0.0126&0.0404&0.0094&0.0063&0.0022&0.0639\\\hline
\end{tabular}